\documentclass[conference]{IEEEtran}
\usepackage{graphicx}
\usepackage[T1]{fontenc}
\usepackage{enumitem}
\usepackage{makecell}
\graphicspath{{./images/}}
\ifCLASSINFOpdf
\else
\fi
\begin{document}

%
\title{DBATES: DataBase of Audio features, Text, and visual Expressions in competitive debate Speeches}

\author{
\IEEEauthorblockN{Taylan K. Sen*, Gazi Naven*, Luke Gerstner, Daryl Bagley, Raiyan Abdul Baten, Wasifur Rahman, Kamrul Hasan, \\Kurtis Haut, Abdullah Mamun, Samiha Samrose, Anne Solbu, R. Eric Barnes, Mark G. Frank, Ehsan Hoque}
}


%


\maketitle

\begin{abstract}
In this work, we present a database of multimodal communication features extracted from debate speeches in the 2019 North American Universities Debate Championships (NAUDC). Feature sets were extracted from the visual (facial expression, gaze, and head pose), audio (PRAAT), and textual (word sentiment and linguistic category) modalities of raw video recordings of competitive collegiate debaters (N=717 6-minute recordings from 140 unique debaters). 
Each speech has an associated competition debate score (range: 67-96) from expert judges as well as competitor demographic and per-round reflection surveys. 
We observe the fully multimodal model performs best in comparison to models trained on various compositions of modalities. We also find that the weights of some features (such as the expression of \textit{joy} and the use of the word "we") change in direction between the aforementioned models. We use these results to highlight the value of a multimodal dataset for studying competitive, collegiate debate.

\end{abstract}

%
\IEEEpeerreviewmaketitle

\section{Introduction}

In the first 1960 United States presidential debate between John Kennedy and Richard Nixon, initial analysis suggested that the radio audience predominantly found that Nixon won the debate, while the television audience found that Kennedy won \cite{gunderman_2016, garsten_2016}. Could their use of facial expressions during the debate help explain this? Several studies have established that nonverbal communication, including facial expressions, pose, and speech audio characteristics, often account for a substantial portion of the meaning conveyed \cite{philpott1983relative, henry2012association, colcta2010importance}. Yet, a fundamental question remains: \textit{How are the different modalities of communication, including textual, auditory, and visual, interdependent in affecting a communication's effectiveness?} 
In this paper, we present a multimodal, expert-labeled, debate dataset for public release that includes several nonverbal communication data features that have often been omitted from prior studies. We then use this dataset to show how focusing on a single modality of interpersonal communication in isolation (as opposed to considering multiple modalities together), often leads one to develop opposite conclusions as to a feature's association with debate performance. For example, we show that when considering facial expressions alone, one is led to believe that smiling with both mouth and eyes (which is often, but not always, associated with joy) are negatively associated with debate score. However, when considered together with the context of textual word category, sentiment, and speech audio features, smiling with both mouth and eyes is shown to be positively associated with debate score. We present these findings along with several others through examination of the multimodal dataset DBATES: DataBase of Audio features, Text, and facial Expressions in Speeches.
\begin{figure}[t]
    \centering
    \includegraphics[width=3.49in]{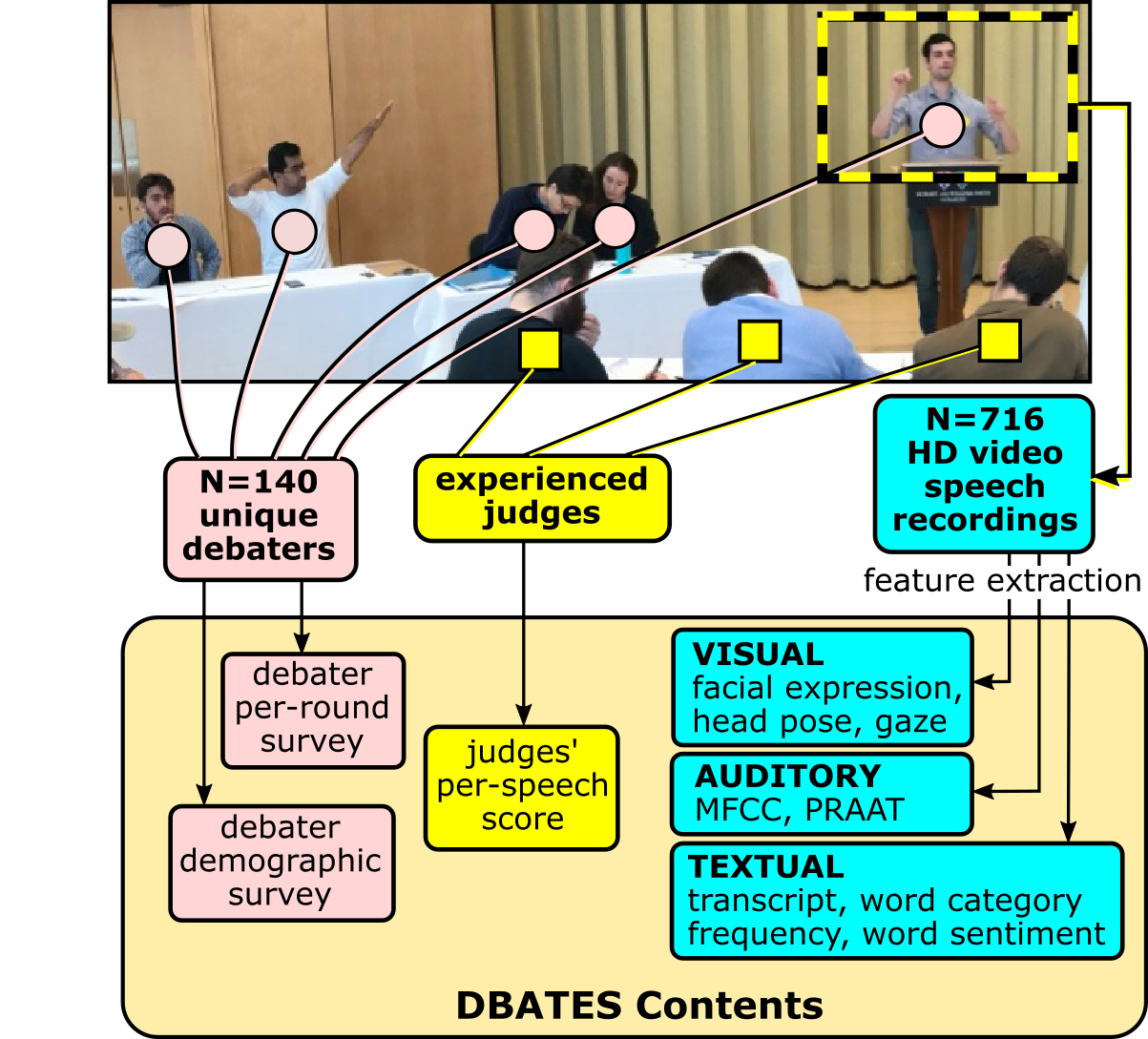}
    \caption{\textbf{DBATES Data Collection and Database Contents.}
    Here we show a portion of a live debate scene at the North American Universities Debate Championships (NAUDC) at Hobart and William Smith Colleges from which the DBATES dataset is based. The yellow and black rectangular box represents the camera view of recordings from which communication features have been extracted. Extracted features include textual, auditory and visual modalities. Each speech also has an associated judge's score and debater survey. 
    }
    \label{fig:flowchart}
\end{figure}

The importance of studying debate goes beyond predicting the outcomes of presidential elections. Studies have shown that education and practice in debate improves one's ability to think critically. For instance, Green and Klug conducted an experimental study to find how debates promoted the quality of critical thinking \cite{TeachingDebates}. They found that students who learned issue through debate had significantly larger increases in critical thinking test performance compared to students in the control group \cite{TeachingDebates}.  Another study demonstrated how debate encourages open thought \cite{KennedyInclass}. More specifically, in their study, Kennedy, et al., showed that through a series of in-class debates, 31{\%} to 58{\%} of participants changed their views after participation \cite{KennedyInclass}. This suggests that students were capable of learning new ideas through debate and recognized merit to different viewpoints to the extent that some ended up adopting alternative views. These studies provide evidence that debates play an integral role in learning new and different perspectives. Perhaps, it is not surprising that the majority of policy makers in the three branches of U.S. government, as well as many world leaders and architects of social change including Nelson Mandela and Dr. Martin Luther King, were debaters in school \cite{burek2014debate}. While these studies make evident the importance of studying debate is evident, the verbal and nonverbal factors which make a good debater are not as clear.

Of benefit to the study of the characteristics of effective debate, several groups have made public debate-related datasets (see Table \ref{tab:table1l}). These debate datasets are fairly recent, and have not previously included combined visual, audio and textual modalities. The Internet Argument Corpus (IAC) released in 2012 contains 390,704 posts from 11,800 discussions, sourced from 4forums.com, an online debate forum \cite{walker2012corpus}. This dataset contains a subset of 103,206 posts and offers a diverse set of labeled annotations across different metrics. While the IAC provides a relatively large sample set, it is not focused on debate nor does it involve modalities other than text. 

Over the past eight years, IBM has been working on "Project Debater" which includes a number of textual and audio samples in the study of debate \cite{orbach2019dataset}. While Project Debater has a number of datasets with over 800 speeches, this project does not include any visual data.

In 2018 Zhang et. al released ArgRewrite \cite{zhang2017corpus}, an argumentative writing dataset that contains 180 essays with custom content and surface annotations. 

Unfortunately, the above corpora of debate-related datasets has a complete absence of any information regarding the visual modality, e.g. facial expressions, head pose, eye gaze, or body gestures. Additionally, many of the mentioned datasets lack a high-stakes, competitive atmosphere. Others lack debate performance labels from high-quality experienced evaluators.

The DBATES dataset vitally addresses these limitations by providing a multimodal, expert labelled, database of competitive debate speeches. DBATES fundamentally includes the previously elusive visual modality, including facial expressions and head pose data, which our analyses show are fundamental in order to properly interpret textual features. The DBATES dataset thus enables exploring the subtleties of debate by observing the interdependencies that are exclusive to a rich, multimodal dataset.

\begin{table}[t!]
    \centering
    \caption{Existing Datasets}
    \begin{tabular}{c|c|c|c}
    \hline
        Year & Dataset & Datapoints & Modalities \\
        \hline \hline
        2012 & \makecell{Internet Arg. \\Corpus} & \makecell{390k \\ Debate Posts} & Text \\
        \hline
        2018 & ArgRewrite & \makecell{180 \\Arg. Essays} & Text \\
        \hline
        2012- & IBM-Rank-30k & \makecell{30k \\Arg. Elements} & Text \\
        \hline
        2012- & IBMPairs & \makecell{9.1k \\Arg. Pairs} & Text \\
        \hline
        2012- & IBM-Debater & \makecell{800 \\Speeches} & Audio, Text \\
        \hline
        2017 & \makecell{Recorded \\Debating \\Dataset} & \makecell{60 \\Speeches} & Audio, Text \\
        \hline
        \hline
        \textbf{2020} & \textbf{DBATES} & \makecell{\textbf{716}\\ \textbf{speeches}} & \textbf{Audio, Text, Visual} \\
        \hline
    \end{tabular}
    \\[5pt]
    The previous premiere datasets on debate in comparison to the DBATES dataset; no prior dataset has included the visual modality (e.g. facial expressions)
    \label{tab:table1l}
\end{table}

In summary, despite the importance of debate, there is currently a lack of debate datasets that allow for the study of facial expressions and gestures in a multimodal manner. In this paper we address these issues, and our major contributions include:
\begin{itemize} 
\item providing the DBATES dataset -  DataBase of Audio features, Text, and visual Expressions in Speeches from the collegiate North American Universities Debate Championships (NAUDC), 
\item finding first order associations and inter-relations of multimodal debate features with expert-judged performance, and
\item crucially identifying that it is necessary to consider audio, facial expressions, and textual features simultaneously to avoid incorrect interpretations in the association of each feature with debate score.
\end{itemize}

\section{Methods}

\subsection{Raw Data \& Collection}

Data was gathered from the 2019 North American Universities Debate Championship (NAUDC) at Hobart and Williams Smith Colleges. This competition was held over three days and involved a total of 224 students, of which 140 participated in our study (i.e., the number of unique participants that we have at least one speech from).

\subsubsection{Recruitment}
Ethical approval was obtained from our university Institutional Review Board (IRB) prior to any participant recruitment. Individuals were recruited through email and electronic flyers which were sent to all tournament registrants. Consent of all participants was obtained prior to any recording or surveying. In addition to a global participation consent, participants data was not used unless they also provided additional consent on a per-round basis. Individuals were motivated to participate through being offered i) high quality video recordings of their debate speeches, and ii) \$5/debate round for answering a short (<2min) survey.

\begin{figure}[tbh]
    \centering
    \includegraphics[width=3in]{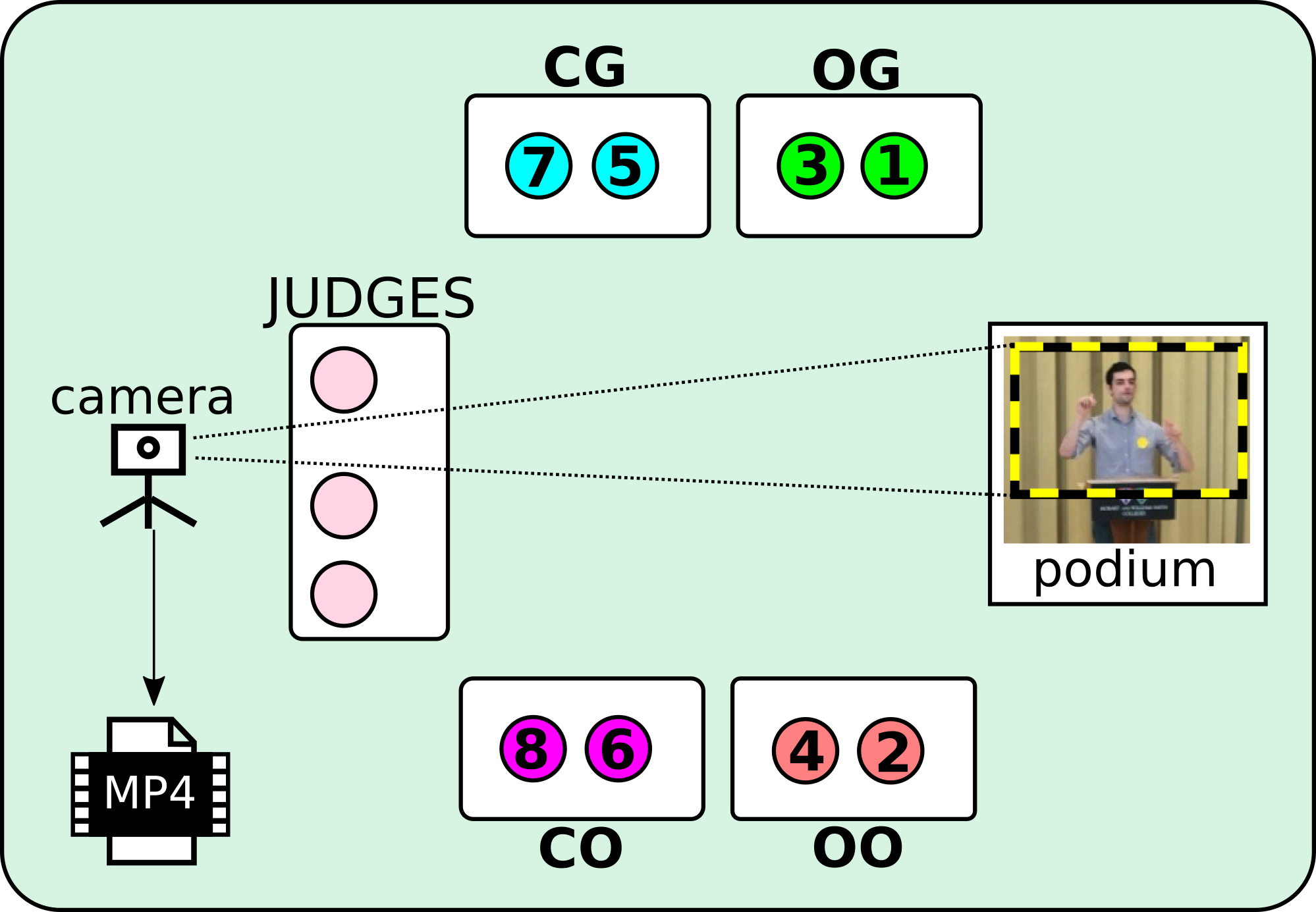}
    \caption{\textbf{Debate Session Room Typical Layout}
    (Seated at the tables are the two-person teams: Opening Government (OG), Opening Opposition (OO), Closing Government (CG), and Closing Opposition (CO), as well as three to seven judges. The numbers indicate the sequence of the speakers in the debate session.}
    \label{fig:scene}
\end{figure}

\subsubsection{Debate Format}
The tournament followed the British parliamentary debate format for varsity level \cite{ecksteinBPdebates}. It consisted of multiple \textit{rounds}, each round had its own \textit{motion} (i.e. issue statement) for teams to debate, and each round lasted approximately one hour. Within a given round, groups of four teams were drawn from all registered teams.  Each group of four was assigned to a given room for their own debate session (as shown in Fig. \ref{fig:scene}). For the first round, selection was completely random. However, for subsequent rounds, point matching was implemented, in which debate teams were grouped with teams that have similar aggregated scores. Each team consisted of two member slots. Within each debate group, two teams were assigned to be \textit{government} (i.e. they were tasked with supporting the given motion) and the other two teams were assigned to be \textit{opposition} (i.e. they were tasked with arguing against the motion). Additionally, the two government teams were randomly assigned to be either \textit{opening} or \textit{closing}. The two opposition teams were likewise assigned to be either opening or closing. The eight debaters in a debate group gave their speech according to the following order. The speaker's designation is :
\begin{enumerate}
  \item Opening Government, ``Prime Minister"
  \item Opening Opposition, ``Leader of Opposition"
  \item Opening Government, ``Deputy Prime Minister"
  \item Opening Opposition, ``Deputy Leader of Opposition"
  \item Closing Government, ``Member of Government"
  \item Closing Opposition, ``Member of Opposition"
  \item Closing Government, ``Government Whip"
  \item Closing Opposition, ``Opposition Whip"
\end{enumerate}
The most common room physical layout of an individual debate group session is shown in Fig. 2. Final and semifinal sessions tookplace in more of an auditorium type of layout. Each speech was limited to seven minutes with an additional 15 seconds of grace time. Speakers had the option of accepting questions from the debaters on the other side of the house.

Although for most teams, each member slot was occupied by a different person, in special circumstances, a single person occupied both slots (aka "maverick"). Thus, each debate group normally consisted of eight competitors.  A debater is considered ``novice" if they are either in their first year of debating at the university level or they have competed in 3 or fewer university level tournaments. If they are not a novice, then they are considered ``varsity." A team is a novice team only if both members are novice debaters. In other words, the level of success at previous tournaments has no impact on novice or varsity status. Only experience is taken into account when assigning these definitions. 

There were 6 preliminary rounds in which all teams competed. Top scoring novice teams were entered into a novice playoff, and the top scoring varsity teams were entered into a varsity playoff. The novice playoff consisted of single elimination semifinals, and finals. The varsity playoff consisted of single elimination quarterfinals, semifinals, and finals.Each motion was secretly selected before the tournament started, by a group of experienced debaters/coaches who were selected as the Chief Adjudicators (CA) for the tournament. Table \ref{tab:motions} in the appendix shows the motions for each round in the tournament. The CA team ensured that the CHAIR judge in every round was an experienced coach or debater. Many of the other judges serving on the panel with the CHAIR judge were also team coaches or experienced debate competition veterans. However, in some cases a few of the panelists were relatively new to the activity. 

For the 2019 NAUDC tournament, each speech was provided a score between 50 and 100. The judges were asked to follow scoring guidelines which emphasize the relevance and number of arguments, strength of the reasoning, clarity, completeness, and vulnerability to rebuttal (see Table \ref{table:rubric} in the supplemental material for complete guidelines). Each debate group of four teams had three to seven judges. While judges were to each provide their own score for each speech, only the average score among the judges was made public. It should be noted there is some disagreement in the debating community about how debaters should be judged. Until recently, the rules stipulated that matter (i.e., argumentative content) and manner (i.e., effective speaking style) should count equally. However, it has been more common practice for years that the judges have increasingly skewed toward deciding based on matter rather than manner . Still, almost everyone acknowledges that the two qualities are not completely independent. Inevitably good manner impacts what judges perceive as good matter. As a result, there are some judges today who believe that manner is currently undervalued in judging these competitive collegiate debates.

\subsubsection{Video Recording}
Sony-HDRCX405 HD Video Recording Handycam Camcorders with optical zoom were used to record speeches at 30 FPS in each of the 30 rooms concurrently where the debate rounds took place. These cameras were used to record both the audio and video of the speakers. The cameras were setup using a tripod for steady recording throughout the speech, and all camera operators were trained to make sure the recording captured the speaker from the top of their head to their waist (See fig 2 yellow and black rectangle and fig 3 example). Each speech was captured with a single recording.

\subsubsection{Demographic Survey}
We developed an online registration application for debaters who were interested in participating in our research study. Through this online portal we also collected information on the debaters age, gender identity, whether they are a native English speaker or not, and their college major.

\subsubsection{Per Round Survey}
After every round of debate, debaters would leave the room and allow the judges to reflect on the round and discuss scores to be given to each debater. During these 15 minutes we released a timed survey, where we asked debaters post round questions. The questions and the possible answers are shown in Table \ref{table:surveyquestion} in the supplementary data. 

\subsection{Extracted Features}
From the raw debate video recordings several data features were extracted including manual text transcription, text sentence sentiment, word category, automated facial expressions, head pose, and speech audio characteristics. Each of these extracted features are provided in the DBATES database.

\subsubsection{Audio Features}
We used Praat \cite{Boersma2009}, an application for analyzing the phonetic properties of recorded speech, to extract audio features from the recordings in our dataset. Each video file was converted into a WAV file, then imported into Praat to extract 14 features including \textit{Mean Pitch (F0)}, \textit{Harmonics to Noise Ratio (HNR)}, and different kinds of \textit{Shimmer (local, apq3, apq5, apq11)} and \textit{Jitter (local, absolute, rap, ppq5)}. Praat calculates each of these features over an automatically adjusted time window. To understand how each feature relates to debate speech score, we calculate the average value of the feature for all windows over an entire speech for initial analysis. In addition, we add standard deviation of pitch as a feature.

\subsubsection{Text Features}
Each recording was transcribed using a pool of professional transcribers. Each transcription contains both the speech text of the debater who has the floor, as well as the text of the questions (``points of information") raised by opposing debaters that the speaker decided to take. The transcripts contain speaker labels, allowing for the removal of the points of information when analyzing the text spoken by the debater.

The positive and negative sentiment of the speech test was extracted using the sentence-level analysis tool provided by VADER \cite{hutto2014vader}. In summary, VADER calculates positive, negative, and neutral sentiment on the sentence-level through the use of a  grammatical and syntactical rule-based model. Thus, VADER is capable of incorporating word order sensitive effects, as well as punctuation as well as slang. Each of the postive, neutral, and negative measures are on a 0 to 1 scale.  Instead of a concrete definition of positve, negative, and neutral sentiment, the features are a collective result of a large number of human raters' understanding of positive, negative, and neutral emotion associated with sentences.  Additionally, VADER provides a compound score which which ranges from -1 to 1. VADER data was analyzed on an aggregated debate speech basis, where the sentence-level features were averaged together by speech and each speech was considered a single datapoint. 

The Linguistic Inquiry and Word Count (LIWC) \cite{pennebaker2007liwc2007} tool was used to analyze word semantic category usage within each speech. LIWC measures the frequency of word category usage in text, in which the word categories were designed, established, and validated through several rounds of evaluation by language experts \cite{pennebaker2001linguistic}. The frequency of word usage in each of the categories has shown to be an effective way of capturing style and in many cases characterizing/classifying high level behaviors. For our analysis, we normalized the LIWC category counts by the total number of words spoken in each debate transcript.

\subsubsection{Facial Expression Features}
The OpenFace tool was used to extract 17 facial action coding system (FACS) \textit{action unit} facial expression levels from the raw video of the debaters \cite{baltruvsaitis2016openface}. The facial action units are provided on a scale from 0 to 5, with 0 representing no expression of the given facial action unit and 5 being the maximum possible intensity of the expression of the given facial action unit. In addition to action units, OpenFace was used to extract eye gaze (x,y) and head pose (6 degree of freedom, i.e. Tx, Ty, Tz, Rx, Ry, Rz).

The Affdex tool was used to extract affective-based facial expression features from the raw debate speech videos \cite{mcduff2016affdex}. Affdex was been trained on over one million facial expression videos that contain rich affective content, and provides expression levels for expressions commonly associated with \textit{joy}, \textit{fear}, \textit{disgust}, \textit{sadness}, \textit{anger}, \textit{surprise}, \textit{contempt}, \textit{valence}, and \textit{engagement} \cite{mcduff2016affdex}. These features were extracted at a rate of 30 frames per second over the course of the speeches. Each of these features are provided on a 0 to 100 scale, with 0 representing no expression of that emotion and 100 representing the maximum possible level of expression (with the exception of valence which is measured from -100 to 100).

\subsection{Analysis Methods}

Several statistical and regression-based analyses were conducted for each of the various feature sets. 

\subsubsection{Statistical Analysis}
In order to identify the differences between high and low scoring debaters, we compared the lower (25th percentile) and upper (75th percentile) quartile scoring groups. 

We chose the Mann-Whitney test, for hypothesis testing, because data distribution for most features were found to be nonparametric. \cite{Neuhauser2011} The hypothesis test were used to determine statistical significance in any differences between the high quartile's median and the low quartile's median for each feature. We also evaluated the Cohen's d effect size and the means of the upper and lower quartiles.
Furthermore, to account for multiple test comparisons, we apply a Bonferroni correction to the significance test results \cite{rice1989sequential}. Differing views exist on how to calculate the Bonferroni multiplier \cite{armstrong2014use, cabin2000bonferroni}. For this paper, we take a conservative approach and treat all features resulting from a single feature group as part of the same hypothesis and thus use a multiplier equal to the number of features in each group.

Additionally for each feature we evaluate the Pearson correlation coefficient  \cite{freedman2007statistics} with the judges' score as well as the correlation between each feature (except where otherwise noted). For the p-values associated with each correlation coefficient Bonferroni correction was applied in the same manner as above.






\subsubsection{Regression Analysis}
To understand the first order relation of the features extracted with the scores of the debaters, linear regression models are trained on each modality separately, as well as on a combined model of with all modalities. Specifically, ridge regression is used (linear regression in which the feature weights are regularized with l2 regularization) and the features and scores are standardized before fit. Different combinations of features from different modalities are reported to show how features behave differently and provide insight on how the model's predictive ability changes. For feature extraction tools that extract features for every frame, such as OpenFace and Affdex, the average over all the frames for a particular speech is computed. Cross validation (10-fold) is used to used to estimate the test and training set errors in predicting the speech scores. The models are evaluated using a mean squared error on both the train and test sets. 
The MSE reported in the subsequent sections are averaged across all of the folds. For each linear regression model, we report the average model weights for each of the features as well. This allows us to understand the importance of each feature in predicting the score, as well as directly see how each of the features are associated with the score in the models.

\section{Results}
The high/low debate score quartile analysis results are shown in (Table \ref{table:ttest}) and correlation analysis is shown in (Table \ref{table:correlation}).
The Mean Squared Error (MSE) for each model is summarized in (Table \ref{table:mse}) and the feature weights of each model is shown in (Fig. \ref{fig:weights}).

    \begin{table}[bth]
    \centering
    \caption{Debate Score High/Low Quartile Comparisons} 
    \begin{tabular}{|l|c|c|c|c|}
        \hline
        Features & \makecell{High \\Quartile\\Mean} & \makecell{Low\\Quartile\\ Mean} & \makecell{Bonf. \\ scaled\\P-value} & \makecell{Effect Size\\(Cohen's \textit{d})} \\
        \hline
        \hline
        LIWC & & & &\\
        \quad \textit{Inclusion} & 0.015 & 0.018 & <0.0001 & 0.42 \\
        \quad \textit{Conjunction} & 0.024 & 0.025 & <0.0001 & 0.28 \\
        \quad \textit{Motion} & 0.0060  & 0.0069 & <0.0001 & 0.34 \\
        \quad \textit{Future Tense} & 0.0028 & 0.0036 & 0.00023 & 0.48 \\
        \quad \textit{We} & 0.0068 & 0.0079 & 0.00091 & 0.31 \\
        \hline
        \hline
        VADER &&&&\\
        \quad positive & 0.11 & 0.12 & 0.020 & 0.30 \\
        \quad negative & 0.059 & 0.049 & <0.0001 & -0.51 \\
        \quad compound & 0.097 & 0.13 & 0.016 & 0.34 \\
        \hline
        \hline
        Affdex &&&&\\
        \quad \textit{Surprise} & 14.8 & 10.1 & <0.0001 & 0.41\\
        \quad \textit{Engagement} & 38.0 & 30.9 & <0.0001 & 0.46\\
        \hline
        \hline
        Openface &&&& \\
        \quad AU01 & 0.37 & 0.33 & 0.0005 & 0.29\\
        \hline
        \hline
        Praat &&&& \\
        \quad F0 - Mean & 220 & 198 & <0.0001 & -0.57 \\
        \quad F0 - SD & 88.2 & 78.8 & <0.0001 & -0.44 \\
        \quad HNR & 4.43 & 4.95 & 0.0006 & 0.39 \\
        \quad Jitter &&&& \\
        \quad\quad local & 0.030 & 0.027 & <0.0001 & -0.73 \\
        \quad\quad rap & 0.017 & 0.015 & <0.0001 & -0.66 \\
        \quad\quad ppq5 & 0.0190 & 0.017 & <0.0001 & -0.76 \\
        \quad Shimmer &&&& \\
        \quad\quad local & 0.19 & 0.19 & 0.00010 & -0.46 \\
        \quad\quad local, dB & 1.68 & 1.65 & <0.0001 & -0.50 \\
        \quad\quad apq3 & 0.090 & 0.088 & 0.028 & -0.31 \\
        \quad\quad apq5 & 0.13 & 0.12 & 0.0064 & -0.36 \\
        \hline
    \end{tabular}
    \label{table:ttest}
    \end{table}  

\subsection{Debate Score High/Low Quartile Comparisons}

Table \ref{table:ttest} lists the features which show a statistically significant difference between medians of the debate score high and low quartiles (using a 0.05 significance level). P-values shown are from the two-tailed Mann-Whitney test, and have been scaled by their associated Bonferroni multiplier for number of features in each modality group. The Effect Size column represents the difference between the quartile means represented in estimated number of standard deviations (effect sizes of 0.2 have been characterized as small, 0.5 as medium, and 0.8 as large.)\cite{cohen1988statistical}

\subsubsection{Text Features (LIWC)}
The LIWC word categories that show a significant difference between the top quartile debaters and the bottom quartile quartile debaters are \textit{Inclusion}, \textit{Conjunction}, \textit{Motion}, \textit{Future Tense}, and \textit{We}. As shown, each of these word categories was used more often on average by the low scoring group than the high scoring group of debaters. The \textit{Inclusion} category includes words in both the \textit{Conjunction} and \textit{We} categories. Thus, it is not surprising that the \textit{Conjunction} and \textit{We} categories have smaller effect sizes than the \textit{Inclusion} category. 

\subsubsection{Text Features (VADER)}
As shown in Table \ref{table:ttest}, speeches given by top debaters contained more negative sentiment and less positive sentiment compared to low scoring debaters. That being said, both groups averaged less negative sentiment than positive sentiment during their debates. Although we omitted showing neutral sentiment in the table due to the fact that there was no significant difference in the amount of neutral sentiment, it should be noted that neutral sentiment was much more common than both positive and negative sentiment for both groups.


\subsubsection{Audio Features}
Vocal audio features including pitch (mean and st. dev.), jitter (local, rap, ppq9), and shimmer (local, local in dB scale, apq3, and apq5) vary between top debaters and bottom debaters with a p-value below 0.05. Two features, Jitter (absolute) and Shimmer (apq11), are omitted from the table because they were not significantly different. The largest differences (in terms of effect size, or in other words, estimated standard deviations) were in the jitter (local, rap, ppq5) with effect sizes ranging from 0.664 to 0.756. The next largest differences are for mean pitch (F0) and average pitch st. dev., both of were significantly higher in the high score quartile. 

\subsubsection{Visual Features}
Overall the top quartile debaters showed a statistically significant higher level of the \textit{surprise} expression feature, with a value of 14.78 
compared to just 10.11 in bottom performing debaters (\textit{d} = 0.41,p-value <0.0001). Similarly, top quartile debaters displayed a higher level of \textit{engagement} with an average level of 38.0 compared to 30.93 for bottom quartile debaters (\textit{d} = 0.46, p-value <0.0001). Among the other remaining emotions they were expressed at a much more similar level between top and bottom debater score quartiles, suggesting that surprise and engagement may be important expressions for debaters to show. 

Table \ref{table:ttest} further shows that the only facial action unit that was statistically significant between top and bottom performing debaters was AU01 (inner brow raiser). Top performing debaters displayed this facial feature with an average value of 0.37 while bottom performing debaters had an average value of 0.33 (\textit{d} = 0.29, p-value 0.0005). The remaining facial action units that OpenFace extracts did not have a statistically significant difference between the top and bottom performing groups.

\subsection{Debate Score to Feature Correlations}

Table \ref{table:correlation} shows the correlations between the various features and the debate scores. Only correlations which were statistically significant at a level of 0.05 are shown. As was done with the quantile analysis, each of the features was averaged separately over each entire speech to create a single multidimensional data point per speech (as scores are assigned to speeches).

    \begin{table}[bth!]
    \centering
    \caption{Feature Correlations with Debate Score} 
    \begin{tabular}{|l|c|c|c|c|}
        \hline
        Features & Correlation & P-value \\
        \hline
        \hline
        LIWC  & & \\
        \quad \textit{Present Tense} & 0.157 & 0.0001\\
        \quad \textit{Swear} & 0.132 & 0.0017\\
        \quad \textit{You} & 0.119 & 0.0050\\
        \quad Not in LIWC dictionary & 0.117 & 0.0053\\
        \quad \textit{Numbers} & 0.106 & 0.0115\\
        \hline
        \hline
        VADER & & \\
        \quad positive & -0.112 & 0.0159\\
        \quad negative & 0.199 & <0.0001\\
        \quad compound & -0.129 & 0.00369\\
        \hline
        \hline
        Affdex && \\
        \quad \textit{Surprise} & 0.212 & <0.0001\\
        \quad \textit{Engagement} & 0.172 & <0.0001\\
        \hline
        \hline
        Openface - AU01 & 0.145 & 0.0029 \\
        \hline
        \hline
        Praat & & \\
        \quad F0 - Mean & 0.202 & <0.0001 \\
        \quad F0 - SD & 0.156 & 0.000704 \\
        \quad HNR & -0.156 & 0.000669 \\
        \quad Jitter (local) & 0.26 & <0.0001 \\
        \quad Jitter (rap) & 0.238 & <0.0001 \\
        \quad Jitter (ppq5) & 0.269 & <0.0001 \\
        \quad Shimmer (local) & 0.17 & 0.000139 \\
        \quad Shimmer (local, dB) & 0.182 & <0.0001 \\
        \quad Shimmer (apq5) & 0.137 & 0.00493 \\
        \hline
    \end{tabular}
    
    \label{table:correlation}
    \end{table}    

\subsubsection{Text Features (LIWC)}
The LIWC categories that showed a significant correlation with score are \textit{Present Tense}, \textit{Swear}, \textit{You}, and \textit{Numbers}.
In addition, normalized usage count of words not in the LIWC dictionary significantly correlates with score. While none of the categories that significantly differ between top and bottom debaters are significantly correlated with score, there is some connection between LIWC word category correlations and top and bottom quartile differences.. \textit{Present Tense} correlates with score, while \textit{Future Tense} is used more often by lower scoring debaters. Similarly, \textit{You} correlates with score, while \textit{We} is used more often by lower scoring debaters. Thus, though the significant categories may not be identical, the findings with these two methods appear to correspond. 

\subsubsection{Text Features (VADER)}
 The VADER correlation results generally align with the results demonstrated by the quartile analysis. As before, the largest correlation of the group is negative sentiment and positive sentiment and a positive compound score are both negatively related to score. Neutral sentiment, which has virtually no correlation, is not statistically significant and is consequently omitted from the table.

\subsubsection{Audio Features}
As shown in Table \ref{table:correlation}, he vocal aspects of pitch, jitter, and shimmer correlate positively with score. The audio features contain the largest correlations our analysis found. Three audio features, Jitter (local, absolute), Shimmer (apq3), and Shimmer (apq11), are omitted from the table because the correlation is not statistically significant. 

\subsubsection{Visual Features (using Spearman correlation)}
From the Affdex tool we find that the \textit{surprise} and \textit{engagement} expression levels have a positive relationship with a debater's score, with correlation values of 0.212 and 0.172 respectively (p-value <0.0001, <0.0001). This suggests that debaters that are more engaged when they are speaking will perform better, and it is also good to display emotions of surprise. From Openface we discover that AU01 (Inner Brow Raiser) has a positive correlation value of 0.145 with the score of debaters (p-value 0.0029). Affdex documentation states that an expression of surprise has an increased likelihood of AU01, and this is found in our data as well.

\subsection{Regression Analysis}
In this section we present the results of the multivariate linear regression models trained on combinations of the different modalities. We specifically show results of how the different modalities interact with each other. The results first exhibit the different mean squared error (MSE) for models using different modalities and then the model weights are presented.

\begin{figure*}[hbt!]
\centering
\includegraphics[width=7.145in]{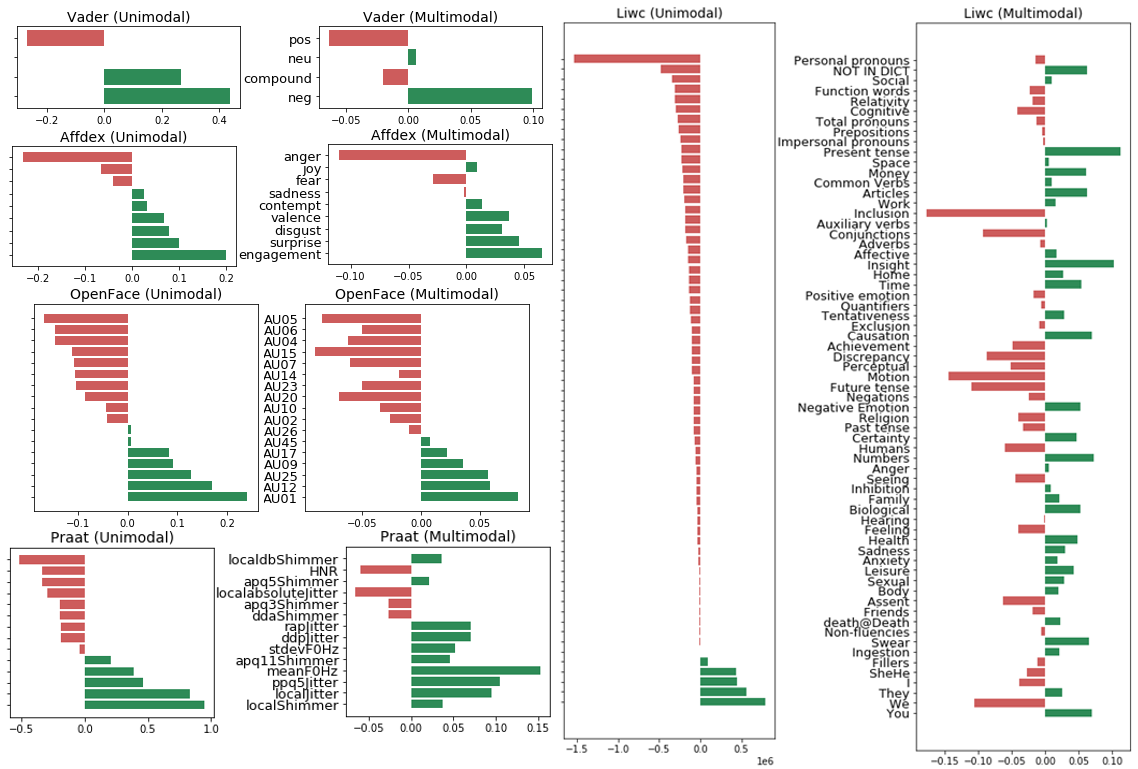}
\caption{Regression weights for unimodal and multimodal models.}
\label{fig:weights}
\end{figure*}

\subsubsection{Mean Squared Error}
    \begin{table}[h!]
    \centering
    \caption{MSE for Linear Regression models} 
    \begin{tabular}{|c|c|c|}
        \hline
        Feature Sets & Train Loss & Test Loss\\
        \hline 
        No features & 7.87 & 8.14\\
        \hline 
        Affdex & 7.71 & 7.94\\
        \hline Vader & 7.83 & 7.91\\
        \hline Openface & 7.43 & 7.78\\
        \hline Praat & 6.82 & 7.07 \\
        \hline LIWC & 6.57 & 7.05 \\
        \hline
       All Features & 5.87 & 6.61\\
        \hline
    \end{tabular}
    
    \label{table:mse}
    \end{table}    
 Table \ref{table:mse} above shows the results from multiple different linear regression models. The first set of linear regression models are using just one specific feature set to predict the debate score. The best model in this case is using the LIWC features, with a train MSE of 6.57 and a test MSE of 7.05. We then combine all of the modalities and this model achieves a train loss of 5.87 and test of 6.61. 
 
\subsubsection{Model Weights}
Figure \ref{fig:weights} shows the coefficient weights for each regression model presented in the previous section. For each feature set that we used we provide a graph that shows the coefficient weights from using that feature set in isolation. We also provide a graph for each feature set showing the weights from using all the features together in a multimodal model. In general we are not considering a comparison between the magnitude of the weights from the isolated models compared to the multimodal models, but rather a comparison of the direction of the coefficient weights. 

Interestingly, there are many feature weights that experience a change in direction in the multimodal model. Specifically the weight for \textit{joy} changes from a negative association with debate score in the isolated Affdex model, to a positive association in the multimodal model. Also, the usage of ``we" and its derivatives were originally a positive weight in the isolated LIWC model, but after considering all modalities the usage of we negatively impacts a debaters score.

\section{Discussion}

\subsection{Multimodal vs. Unimodal Analysis}
Statues and paintings of Lady Justice, the personification of fairness and morality in a judge, often depict her wearing a blindfold and holding a scale. The blindfold represents impartiality, there to free her from visual bias as she uses the scale to accurately weigh conflicting evidence. There are many modern day examples in which one or more sensory modalities is obscured in hopes of obtaining a better judgement. Wine judges are served wine in opaque black wine glasses for some competitions in order to prevent being affected by the color of the wine, music school and orchestra candidates are given "blind" auditions behind a curtain, and even the common adage encourages us to "don't judge a book by its cover". Yet, in several instances, analysis of the DBATES dataset shows that it is necessary to consider nonverbal context in order to properly judge a speech. Specifically, the feature weight disparities in the unimodal and multimodal regression models in the previous section demonstrate how considering one modality in isolation of the others yields one to make incorrect conclusions about a speech feature's association with debate success (i.e. score). One of the features showing such a disparity is the usage frequency of \textit{We} category words (The \textit{We} category words include: we, we'd, we're, we'll, us,  our, ours, lets, let's).

The unimodal regression analysis utilizing only LIWC word category frequencies finds a positive weight association between \textit{We} usage frequency and debate score. Using the LIWC features alone, one may be inclined to deduce that a speech with increased \textit{We} usage is more likely to be associated with a higher debate score. However, as evidenced in the multimodal model, increased \textit{We} word category usage is associated with lower debate score. This demonstrates that among the full set of features across all modalities, there are interdepenencies. Another way of looking at this phenomenon is by considering \textit{We} to be the only dependent variable, while all other features are confounding factors. Because the confounding factors are not distributed evenly among the different debate scores (i.e. Fig. 3), unimodal analysis ends up producing incorrect results. With the benefit of a rich multimodal dataset, we are more likely to discern the true association between each feature and debate score. 

As shown in the Liwc multimodal weight graph in Fig. 3, we see that \textit{We} is negative associated with debate score. While this data does not indicate causality, it does suggest that speakers may be able to improve their debate score by using less \textit{We} language. This is perhaps surprising, as putting the group before oneself is a characteristic often culturally and socially extolled. In the United States the mantra \textit{There is no I in TEAM} is commonly heard in school, sports, and workplace environments.  Prior literature has examined the effectiveness of using ``we" in some of contexts. For example, ``we" can signal a sense of group identity, and previous research has shown that the more a couple used ``we", the better their marriages \cite{simmons2005pronouns}. Tausczik, et al. showed that the pronouns people use often provide useful insights to where their focus is \cite{tausczik2010psychological}.   

This has been shown in the realm of advertising, where positive political ads were shown to use more personal pronouns such as ``we" and ``I" than negative ads, which focused more on the opposition \cite{gunsch2000}. 
Although ``we" has been utilized in various domains successfully, the positive effects of using ``we" are not conclusive and the literature is divided depending on the specific usage case. For example, use of first-person plural pronouns such as ``we" was found to be negatively related to group cohesiveness in groups consisting of 4-6 members \cite{gonzales2010language}. It may be the case that the individuals in this group were using such language to try to create, or give the appearance of, cohesion in an otherwise noncohesive group. We can gain much from understanding how people choose their words and in which situations those word choices are beneficial. With the literature divided on whether or not using ``we" is beneficial, it is not surprising that our analysis shows contradictory results regarding the effectiveness of  \textit{we} depending upon whether one uses a multimodal vs. unimodal approach.

Looking into the various contexts that ``we" is used in competitive debate could yield answers as to what the potential utility of ``we" might be.  In a debate round, ``we" can be used either to refer to a debater and their partner, or can be used to refer to some larger community of people. The former usage is very common in debate. In some cases, it is used merely as a signpost to clearly demarcate what is being asserted by one's own team, in contrast to what others are asserting (e.g., ``They tell you that it is worth trading some liberty for security, but \textit{we} say that this is never a good bargain."). One can imagine a team using this signposting function of ``we" too little. This could render it unclear what claims are being asserted by that team and what claims they are refuting. This could lead to a confusing speech that was unpersuasive, in which case the low frequency of ``we' usage results in a lower score.

In other cases, ``we" functions as a filler word that ultimately waters down what is being asserted (e.g., \textit{We} think that the government has an obligation to protect the most vulnerable in society" as compared to just ``The government has an obligation to protect the most vulnerable in society"). For the same reason English teachers have been telling students to remove the ``I think" (or worse ``I feel") from their essays, it is wise for debaters to remove these uses of ``we" from their speeches. For some debaters, ``we" is largely just an empty filler word, and for others it also waters down the force of their assertions on top of wasting valuable time. In this case, higher frequency of ``we" usage results in a lower score.

In contrast to both these cases, one can imagine other uses of ``we" that are based on trying to include the audience in a broader community of like-minded thinkers. For example, ``Some people may naively embrace neo-liberal capitalism as a pure meritocracy, but \textit{we} all know that this is a terribly simplistic, cruel and ultimately racist perspective." This example sentence attempts to bring the judges into the wise in-crowd being defined by the speaker, potentially causing a higher score to be given for that particular debater using ``we" in that specific case. 

Clearly there are plenty of useful and harmful ways that ``we" usage comes up in debate. The jury is still out on whether a mere frequency count of the word is sufficient to explain its effectiveness or lack thereof. We hypothesize that the context regarding whether \textit{We} usage is good or bad can be explained by the facial expressions and audio features contemporaneous with ``we" word usage. In regards to our dataset from the NAUDC debate tournament, we argue that a rich multimodal model is essential to properly determine the appropriate frequency of ``we" usage in the varying contexts in which it can occur during a competitive debate climate. 

The level of \textit{joy} expression like \textit{We} usage frequency experiences significant change when all modalities are taken into account. When considering the unimodal model, higher levels of \textit{joy} facial expression are associated with low overall debater score. However, when we look to the multimodal model for a more detailed understanding, we find that \textit{joy} expression usage has a positive association with debate score. Fig. \ref{fig:weights} illustrates an example of the the \textit{We} word category and \textit{joy} expression levels may interact in producing the debate score. In the first row a participant from the study used ``we” with 0.019  frequency (average: 0.0076) and showed an average expression of 0.004 for \textit{joy} (average: 1.233), and achieved a score of 70 (bottom quartile of speakers). For this individual if only the LIWC features were considered, their above average use of ``we" would have led to a higher performance prediction. Additionally, if only the Affdex features were considered, a higher score would have been expected due to lower levels of \textit{joy} than average. Furthermore, in the unimodal model for Affdex, lower expressions of \textit{joy} impact the score in a positive way.  However, this debater received a lower than average score. This clear observation demonstrates the importance of utilizing all modalities to gain a better understanding of performance. Another debater as shown in Figure 5 used ``we” with 0.0034 frequency, showed an average expression of 8.57 for \textit{joy}, and achieved a score of 82 (top quartile of speakers). This speaker would have been expected to perform poorly if only the LIWC or Affdex features were considered due to using ``we" less often and having a very high expression of \textit{joy} throughout their speech. However, in the multimodal model the weights for ``we" frequency and \textit{joy} undergo a change in direction, and a more accurate prediction for this debater's score is achieved.

\begin{figure}[ht]
    \centering
    \includegraphics[]{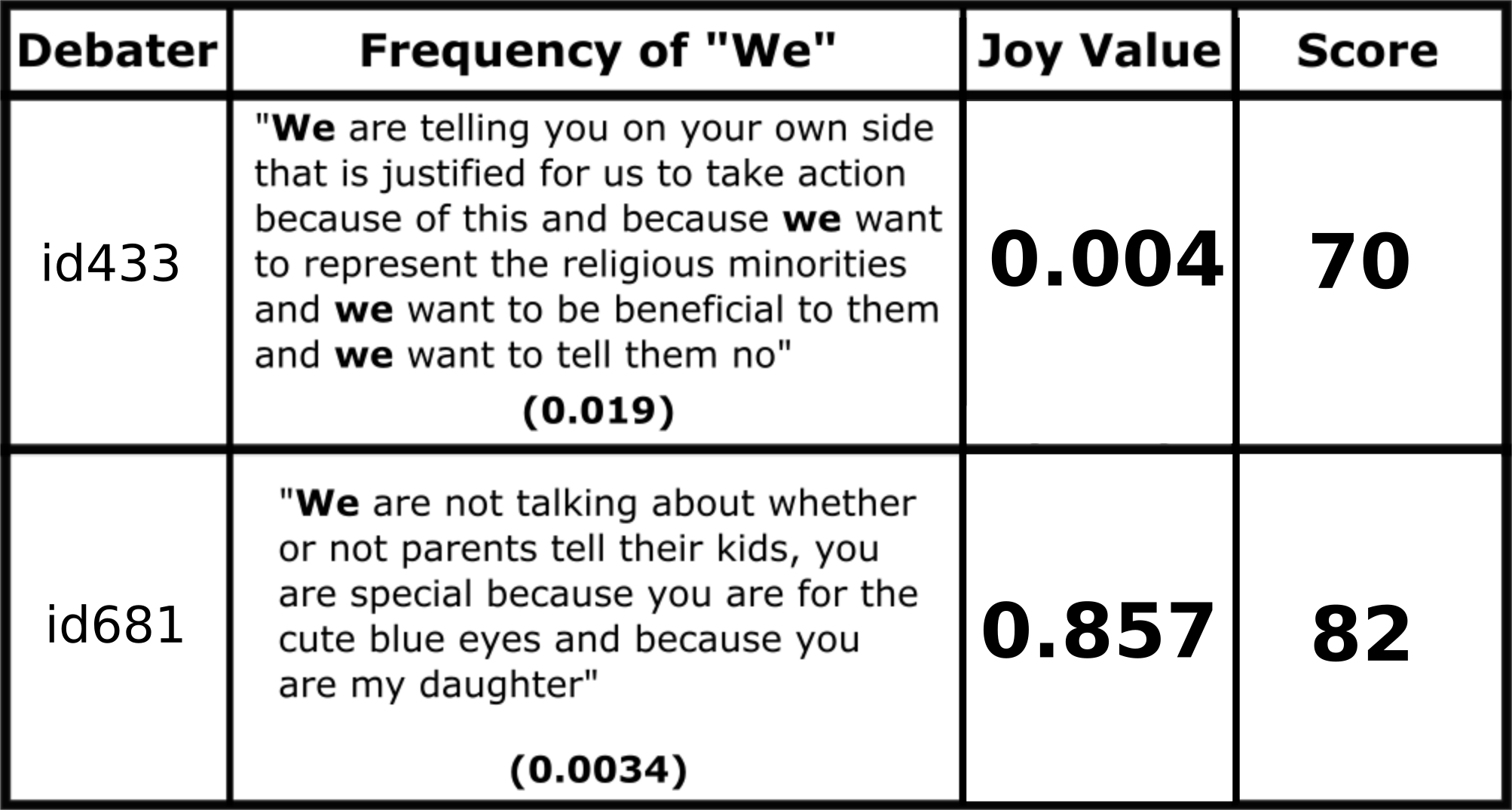}
    \caption{\textbf{Top debaters use ``we" less frequently and have higher levels of \textit{joy} expression}}
    \label{table:wejoy}
\end{figure}

We are confident our work demonstrates the superiority of a modally comprehensive analysis in its ability to more accurately model the existing inter-dependencies contributing to the scoring of competitive debates and provides insights into some of the important variables regarding such a task.


\subsection{Feature Associations with Debate Score}
\subsubsection{Textual Modality - LIWC}
Whereas the LIWC \textit{We} word category experienced a change in interpretation between unimodal and multimodal analysis, several word categories exhibited significant assocations with debate score which were not changed between the different types of analyses including the hi/low debate score quartile analysis, the unimodal regression analysis, and the multimodal regression analysis. The frequency of \textit{Conjunction} words (e.g. or, and, but, also, although, unless, however, etc.) was lower in the higher performing debaters (p<0.0001 in quartile analysis, and the multimodal regression model finds \textit{Conjunction} to have the 5th most negative weight out of 108 features). Perhaps better debaters use more concise and concrete language, and less run on sentences. This may make their arguments easier to understand. 

Similarly, the lower use of the \textit{Future tense} word category in stronger debaters may indicate increased use of real-world data rather than hypothetical projections (p<0.0003 in quartile analysis, and the multimodal regression model finds \textit{Future tense} to have the 3th most negative weight out of 108 features). In a similar vein, the strongest correlation with debate score was found with the \textit{Present tense} word category. The strongest statistical significance, as well as the second strongest effect size, is seen in the frequency in the use of words in the \textit{Inclusion} category. This is likely due to the fact that the \textit{Inclusion} category includes words in both the \textit{Conjunction} and \textit{We} categories. 

\subsubsection{Textual Modality - VADER}
The quartile and regression results (both multimodal and unimodal) also show that higher scoring debaters tend towards using more negative sentiment and less positive sentiment compared to lower scoring debaters  (p<0.0001, p<0.02). This suggests that choosing to use more words with a negative connotation and less words with a positive connotation leads to arguments appearing more persuasive, possibly by sounding more draconian. Perhaps in the context of debate, by using a negative sentiment, a debater can frame their argument in order to rally support to the negative element that needs to be changed. Similar results have been demonstrated in other domains. For example, evoking negative emotion has been shown to be more effective during fundraising activities \cite{Fisher2008}.  

\subsubsection{Visual Modality - Affdex and OpenFace}
Through quartile, correlation, and regression analysis we also found that both surprise and engagement expressions were significantly different between the top and bottom performing debaters (all p<0.0001). The difference in \textit{engagement} rather obviously suggests that being more engaged with the judges (the judges were located directly in front of the camera) helps a debater. Furthermore, emotion expressions are important to consider when modeling debate speaker score because evidence has been shown that emotion plays a large role in how speeches are perceived. Gonzalez et al. showed when a leader speaks using a great amount of emotion it may be transferred to the audience and can increase the level of support for that leader \cite{gonzalez2012emotions}. A similar pattern may be happening in these debate speeches. When debaters are displaying more engagement and surprise, the judges take notice of this and it may influence higher scoring for that debater. 

Overall AU01 (inner brow raiser) was the only OpenFace feature statistically different between the top and bottom performing debaters, with top performing debaters expressing higher average levels of AU01 (p<0.0005). This along with the regression weights in Fig. \ref{fig:weights} show that having more expression delivered through the eyebrows and not having a static upper face region can positively impact the debater's score. Anecdotal accounts of AU01 action have been seen as an expression of conviction, and if this is how it appears to the judges, then that might explain why it would be more prevalent in the high scorers. It is possible, given that Affdex analysis indicated showing more \textit{surprise} is associated with the top-tier debaters, that AU1+2 would also be significant (as AU1+2 is included in \textit{surprise}). However, our initial analysis on action unit level intensities was limited to individual action units only, as we relied on Affdex to understand more complex combinations of facial expressions.

While expression levels of \textit{Joy} were not significantly different between top and bottom performing debaters, our multimodal model found a small positive association between \textit{Joy} levels and debate score. Our findings from OpenFace in that there was no statistically significant difference between AU06 or AU12 between the top and bottom quartile debaters. It is surprising that there is no a stronger association between smiling and debate score, as smiling has been shown to be an important expression to show while speaking \cite{tartter1980happy}.  Perhaps given the combative nature of debate, smiling frequently is ill advised. It is possible that expressions of joy were associated with a decreased perception of seriousness, and thus not as likely to have a strong positive effect on debate score as \cite{gonzalez2012emotions} and \cite{tartter1980happy} would predict. This may especially be true due to the serious nature of the topics debated, including religion, climate change, military interference in Syria, and childcare (see Supplementary data Table V). Overall, as shown by the significance of expressing \textit{Engagement} and \textit{Surprise} expressions, the importance of not remaining in a neutral face is clear. 

\subsubsection{Audio Modality - Praat}
We found that several audio features (see tables \ref{table:ttest} and \ref{table:correlation}) were statistically significant between high scorers and low scorers, and that many of those features were statistically significantly correlated with score. This suggests that higher scoring debaters speak differently than lower scoring debaters based on a unimodal analysis of the audio modality. 

Given that judges are expected to score debaters based on the quality of their arguments, rather than the quality of their presentation, one might expect to see little to no correlation between other features and score (except where such features correlate to the quality of the argument). Consequently, the relatively high effect of audio features on score is surprising. Perhaps this suggests a connection between how a speaker expresses their points and the quality of the points they suggest through a mechanism like confidence or pacing. Alternatively, this could be evidence of another confounding factor that happens to impact both audio and score, such as gender, posture, health, or stature. Many participants declined to specify their gender and in order to preserve the most data for the multimodal analysis, gender was not included. Of the participants that reported male and female gender, there was no statistically significant difference in mean score (F=76.41,M=76.43, std.dev.=2.80, N=580).

\subsection{Nonverbal Immediacy}
The concept of presenters or teachers demonstrating nonverbal immediacy \cite{mehrabian1971silent} has been shown to correlate significantly with learning outcomes across cultures \cite{pogue2006effect} as well as student motivation \cite{christophel1995test}. 
Specifically, nonverbal immediacy consists of using body and facial gestures, smiles, more body lean toward the class, and non-monotonic speech \cite{matsumoto2013emotional}. 
It appears in the multimodal analyses that \textit{engagement} may capture some of that concept, but we see other variables capturing other elements, such as the variability of voice tone (standard deviation of fundamental frequency in Fig. \ref{fig:weights}), and maybe as well the use of AU01, which is often paired with AU02 to generate eyebrow raises – which are in fact facial gestures. Thus, this data strongly suggests the more immediate debater is scored higher.

\subsection{Regression Model Accuracy}
The all-feature (multimodal) regression model had a mean squared error of 6.61, compared to 8.14 for the featureless model, showing that our model is able to explain 18\% of the debate score variation. Given that the features are averaged over the entire duration of the speech, it is perhaps surprising that the features were able to provide any improvement over the featureless model in predicting debate score. In order to better predict debate score, it is likely necessary to understand the deeper meaning of each of the arguments that comprise a given speech. Additionally, as table IV shows that the error was still substantial for the training loss, it is clear that a model with more complexity than a linear model is necessary to learn the training data. In future work, we are hopeful that more sophisticated models will be developed to explain this dataset more completely. However, linear models are useful in that their interpretability allows the general trend to be understood in the association of the features with the dependent variable; debate scores. From Table \ref{table:mse} it is apparent that as features from different modalities were added, the mean squared error loss of the linear regression decreased. This is not surprising given our rationale for advocating multimodal modeling because with each additional modality comes more information. The additional information enables the model to be more certain of its predictions and this observation provides concrete evidence of the need for analyzing debates using multimodal data. Interestingly, unimodal feature sets such as Praat and LIWC turned out to be relatively more accurate than their unimodal visual counterparts (this is likely because predicting debate score based on AU intensity-level patterns has its limitations given that judges are instructed to focus on the words of the argument rather than the speaking style). However, the train and test error for the unimodal Praat and LIWC models are still higher than that of the multimodal model with all features. Thus, there is evidence suggesting that there exists information present in the visual modality that is missing from the others.


\subsection{Limitations}

\subsubsection{Application to the first Kennedy-Nixon Debate}
\begin{figure}[tbh]
    \centering
    \includegraphics[width=3.49in]{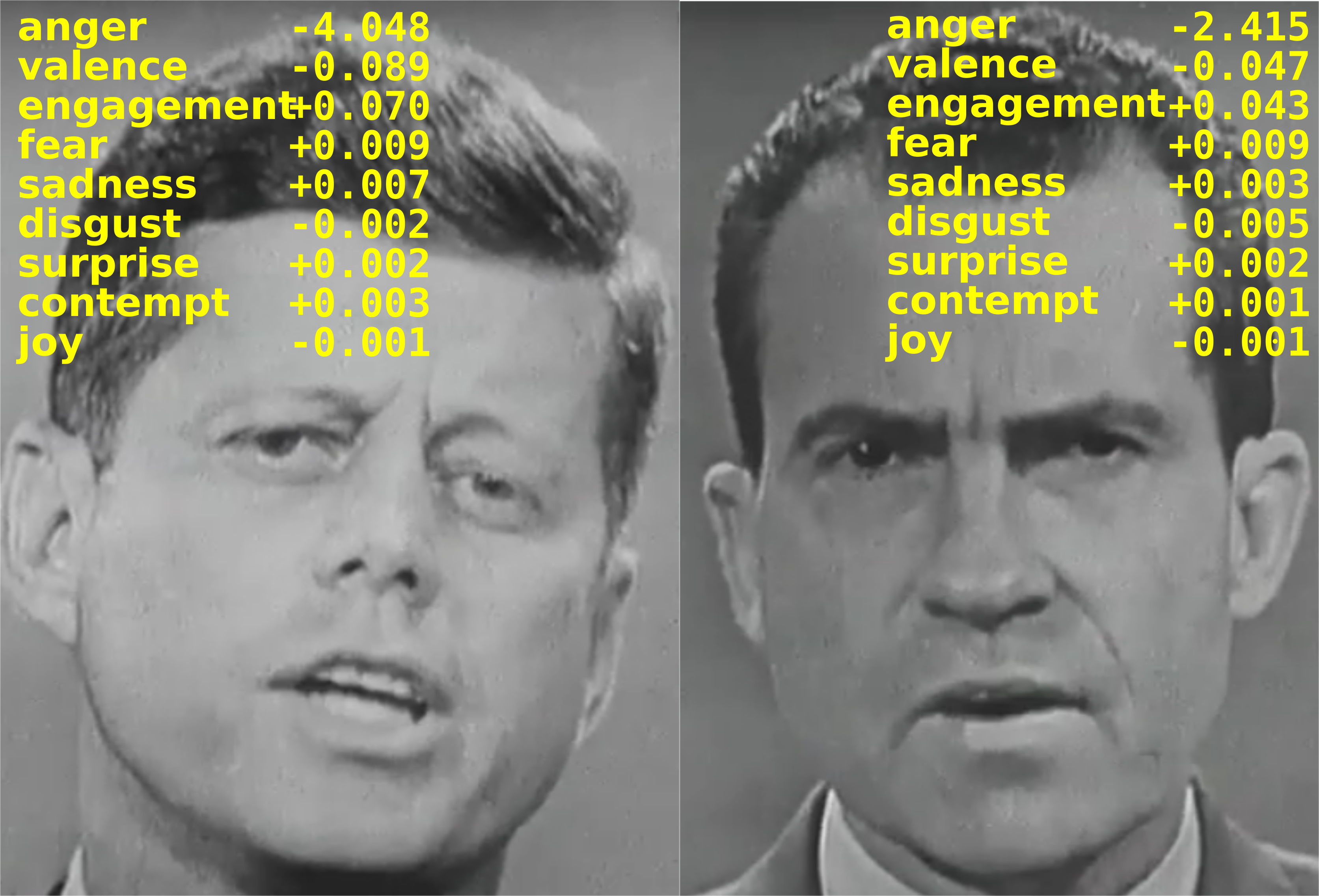}
    \caption{\textbf{Predicted effect on debate score by averaged Affdex visual features on the Kennedy - Nixon debate speeches.}}
    \label{fig:jfk_nixon}
\end{figure}

In the introduction, we mentioned that in the first presidential debate from Kennedy and Nixon in 1961, the radio audience found Nixon to have won, while the television audience found Kennedy to win \cite{gunderman_2016, garsten_2016}. This interpretation, however, is controversial, with alternative analysis suggesting that there were no substantial disparity between television and radio audiences \cite{campbell2016getting}. As an anecdotal example of applying the DBATES multimodal model to a real world example, as well as to help demonstrate model limitations, we compare how the model predicts the Kennedy and Nixon debate speeches with and without visual data. Shown in Fig. \ref{fig:jfk_nixon} are the predicted effects that the automatically extracted emotion expression levels have on debate score. The model predicts that both Kennedy's and Nixon's \textit{debate score} would be significantly reduced by each candidates high \textit{anger} (and low \textit{valence}) expression levels, with Kennedy's reduction greater (4.048 score reduction for Kennedy and 2.415 reduction for Nixon). The model also predicts that Kennedy has higher expression levels of \textit{engagement} compared to Nixon, which would give Kennedy 0.027 score advantage over Nixon. The effects of the other visual features are in comparison minimal. In summary, the model would suggest that Nixon would benefit more from the visual modality conveyed in television. However, it is important to note that the Affdex tool predicted average \textit{anger} expression levels of 55.9 and 33.5 for JFK and Nixon respectively. The training data involving tournament competitors displayed no where near this high level of anger, with the largest average anger level detected being 11.2. Perhaps Affdex is over-estimating the perceived anger levels due to the prominent brow furrows displayed by both Kennedy and Nixon (see Fig. \ref{fig:jfk_nixon}. Because our model is multimodal and dependent upon a large number of features, its estimation of debate score becomes more diversified and less susceptible to adverse performance due to a single feature being off. The large disparity in anger levels of JFK and Nixon brings light to a limitation of the DBATES dataset in that real world data may exhibit feature input feature values outside of the range of values found in the dataset. While a collegiate population is generally diverse in many aspects, the fact that the level of \textit{anger} expression detected in the JFK - Nixon debate speeches fell outside the DBATES range raises caution in its application to other real world scenarios.

\subsubsection{Expert Judge Scoring}
Although the DBATES dataset possesses the unique qualities of being the first debate dataset of its kind to include the visual modality and the dataset is expertly labeled, there is one potential limitation of these two qualities that should be mentioned. The debater score that is given by these judges may not accurately incorporate the visual modality into its assessment in a way that a lay person would. This is due to the observation that the scoring rubric the judges are given stresses matter (i.e., argumentative content) over manner (i.e., effective speaking style). It is reasonable to assume that due to the scoring guidelines, some judges would have the incentive to not look at the speaker throughout their debate and instead focus on listening to the argumentative content while taking notes. It would directly follow for that particular judge giving that specific score for that instance of speaker, that the visual modality is not reflected in their overall score. In other words, if a judge isn't looking at the speaker, then the speaker's facial expressions or other body language presumably cannot affect the judge's evaluation of that speaker. However, we argue that most judges do in fact look at the given speaker and that it is nearly impossible to completely ignore the influence of manner on debate quality even when explicitly trying to do so.

\subsubsection{Feature Analysis}
A limitation of the LIWC tool specifically is the low dictionary size of less than 6000 words/word roots, which represents only 6.2\% of the debate transcript corpus overall. Automated facial expression analysis software has and continues to experience rapid advance in its ability to accurately detect facial landmarks \cite{baltruvsaitis2016openface}. However, there has also been recent criticism on the universality of facial expressions associated with truly experienced emotion. We would like to stress that the Affdex facial expression analysis is directly related to the expressed emotion, and our findings are not dependent upon an equivalence between actual and expressed emotion. Some of the auditory features (such as mean F0 pitch) may benefit from the use of gender labels. Our analysis does not make use of self reported gender since several participants did not report their gender, but we foresee future alternative analyses of the DBATES database investigating the use of demographic variables.

\subsubsection{Data Collection Procedure}
While pursuing to collect as much data as possible, inevitably, we encountered limitations during the data collection process. To protect the privacy of the speakers, we only recorded individuals that opted into being recorded. This means that for some debates we may not have all eight participants recorded.
 Another precaution we took to protect the privacy of the participants was agreeing to not release the raw audio or videos of the individuals. Instead, we have opted to release commonly analyzed features (such as Affdex and OpenFace AUs) and run any other analysis on the raw data by request. Indeed, not having the option to view the raw video footage could prevent researchers from drawing conclusions intuitively based on watching the videos.

In striving to record as many debaters as possible, we were required to enlist a large research staff to record debates because they happen simultaneously. This leads to potential inconsistencies in the way the videos were recorded, as well as in how easy it is for the audio to be transcribed. Consequently, the transcribers may have made errors or marked a portion "inaudible".

Finally, this paper focuses on the dataset and does only basic analysis. The machine learning models we used to understand the data are therefore simpler than one could use to try to best predict debater score (or another quality). However, by sacrificing performance scores, we acquire far more interpretability from employing said models.

\subsection{Future Work}
There are many exciting directions for this research moving forward. We have self-reported surveys on a debater's level of belief in the position they were arguing for. Using that information, we could work to develop a model that helps detect when a person is voicing a position that they do not actually hold. Such information could be valuable at a grand scale, such as for evaluating the speeches of foreign government officials (e.g., understanding whether a public speech is meant to inform or to propagandize). Such information can be applied to understand small scale phenomenon as well, such as detecting specific subtleties of human communication (e.g. sarcasm and insincere flattery). Machines that can better recognize the subtle nuances of human communication will be able to better serve users in a variety of contexts.

Another interesting idea is to develop a system where an individual can practice giving a speech/debating a topic. We could apply the insights learned from deploying that system to help people mediate their political disputes over social media platforms. Imagine if there were a software add-on that would assist in providing objective feedback such as identifying toxic exchanges, checking factual claims or emphasizing forgotten points. Given the current opinion wars being waged on social media, any tool that could enable a more productive digital discourse would be highly sought after on both sides of the house.

This research could be useful in quantifying a ‘gut’ impression. Often, people make ‘gut’ judgments – and while some have terrific track records making such judgments, others have terrible track records. One effort on ‘gut’ judges are the ‘wizards’ of deception detection who seem to routinely get 80\% or better detecting deception \cite{o200412}. They often cannot articulate why they believe someone is lying, but yet they have shown consistent accuracy. Multimodal analysis methods may help detect those elements that make someone an expert. For example, interviews with horse race handicappers show these individuals use a six factor function when rating horses, and this cognitive skill is uncorrelated with IQ \cite{ceci1986day}. This decomposition was based on interviews and then correlated with publicly available information on each horse (closing speed, weight, weather, etc). The analysis methods discussed in this paper present a potentially big step in further decomposing how such individuals process their worlds.

Another area that would be interesting to pursue would be to have the debater speeches evaluated by third party non-experts (e.g., Amazon Mechanical Turkers). Barnes et al showed that there is a clear difference between experienced and lay judges in the realm of evaluating debate \cite{Barnes2015}. It would be interesting to compare the ratings from the non-experts with the experts and identify what features are contributing to the discrepancy of the ratings.  This would have important ramifications for presidential elections specifically because most of the voters would be classified as non-experts for evaluating presidential speeches/debates.  

In addition to the data planned for the initial release, we have more data on the self-reported surveys further increasing the richness of the dataset. In addition to the measures of how strongly a debater felt in support of the topic they were debating, we captured whether the debater was a novice or varsity and even had the participant rank the teams from first to last from their perspective. We also have data on how many times various teams were referenced by other teams in the forms of rebuttals and points of information. Although the video data needs to be sifted through manually and annotated, we are excited about the potential of including reference counts for predicting debater score. Looking into all these factors will be a very exciting frontier of future debate research.

\section{Conclusion}
Debate is a useful skill for explaining and exploring ideas for individuals and societies. Nevertheless, up to this point there has not existed a debate dataset that includes visual data. To fill that gap, this paper presents a novel, multimodal debate dataset containing over 700 unique speeches. From this dataset we are able to analyze facial expressions, affect, phonetics, text sentiment, and LIWC categories to see how those features relate to participant surveys and official judge scores.

Furthermore, this paper demonstrates how having multiple modalities allows for a greater understanding of debate scores and how to foster more accurate predictions. Naturally, adding more information leads to higher accuracy and lower error rates. In addition to that, by analyzing the various modalities both individually and simultaneously, we showed that some features, such as use of the word ``we" and facial expressions of joy, change in meaning when considered in light of all modalities. This suggests that features can be misinterpreted when modalities are missing and better understood when taken in the context of modally diverse feature sets. Consequently, we believe that this dataset is a valuable resource for the continued study of debate and multimodal machine learning models and therefore encourage other researchers to use the dataset for future work.

\section*{Acknowledgment}

This research was supported in part by grant W911NF-15-1-0542 and W911NF-19-1-0029 with the US  Defense Advanced Research Projects Agency (DARPA) and the Army Research Office (ARO), and the National Science Foundation NRT-DESE \#1449828. The authors would like to thank the HWS debate team for their help. The first two authors should be considered co-first authors.



\bibliographystyle{IEEEtran}
\bibliography{IEEEabrv,ref}
%



\begin{IEEEbiography}[{\includegraphics[width=1in,height=1.25in,clip,keepaspectratio]{images/Taylan.jpg}}]{Taylan Sen}
Taylan Sen received the BS degree in Electrical Engineering and MS degree in Biological Engineering from Cornell University and the JD degree from University at Buffalo. He has seven years industry experience as a software engineer and five years experience as an intellectual property attorney. He is currently working toward the PhD degree in Computer Science at University of Rochester. His research focuses on computational models of nonverbal communication and is conducted under the supervision of Prof. Ehsan Hoque.
\end{IEEEbiography}

\begin{IEEEbiography}[{\includegraphics[width=1in,height=1.25in,clip,keepaspectratio]{images/Ehsan.png}}]{Ehsan Hoque}
Ehsan Hoque received his Ph.D. degree from the Massachusetts Institute of Technology in 2013. He is an associate professor of computer science with the University of Rochester where he co-leads the ROC HCI Group. Hoque’s research aims to use techniques from artificial intelligence to amplify human ability. His research has been recognized with the MIT TR35 Award, NSF CAREER Award, ECASE-Army Award, among others. He is a member of the ACM, IEEE and AAAI.
\end{IEEEbiography}

\clearpage
\section{SUPPLEMETARY DATA}

Some tables referenced in the paper provide additional information not necessary to explain the contributions we present, but still provide information the readers may find valuable. Such tables are included here.

\begin{table}[h]
    \caption{Motions for Each Round in the Tournament (Rounds 7: Novice Semifinals, 8: Novice Finals, 9: Experienced Quarterfinals, 10: Experienced Semifinals, 11: Experienced Finals)}
    \centering
    \begin{tabular}{|c|c|}
    \hline
        Round & Motion \\
    \hline
        1 & This house will punitively tax personal homeownership. \\
    \hline
         & This house, given limited resources, will  prioritize \\
        2 & funding for early childhood care to the exclusion \\ 
         & of funding of post-secondary education.\\
    \hline
        & This house regrets the narrative that ideal political\\
        3 &  debate should be objective to the exclusion of\\
        & emotions and subjective experiences.\\
    \hline
        4 & This house believes that ASEAN [Association of Southeast \\
        & Asian Nations] should adopt a common currency.\\
    \hline
        & This house believes that federal governments should  \\
        5 & suspend all funding to sub-national governments  \\
        &  (states in the USA and Mexico, provinces in Canada) \\
        & that violate religious liberties.\\
    \hline
        & This house believes that parents should teach their \\
        6  & children that they are inherently special (e.g., in what  \\
        & they deserve, and/or what they are capable of achieving).\\
    \hline
        7 & This House supports the use of truth and reconciliation\\
        & commissions over courts in post-conflict nations.\\
    \hline  
        8 & This house prefers a polytheistic society to a \\
        & monotheistic one.\\
    \hline
         & This House, as Russia, would announce that any \\
        9 &  further Israeli strikes in Syria will be met by a \\
        & Russian attack on Israel. \\
    \hline
        & This House wishes to allocate resources towards  \\
        10  & adapting to climate change, rather than mitigating/ \\
        & preventing climate change.\\
    \hline  
        11 & This house believes that it is morally legitimate  \\
        & to avoid taxes in undemocratic regimes.\\
    \hline
        
    \end{tabular}

    \label{tab:motions}
\end{table}{}

\begin{table}[h]
    \centering
    \caption{Questions asked on the survey after each round}
    \begin{tabular}{|c|c|}
    \hline
        Questions & Answer Choice\\
    \hline
        To what extent do you personally & Strongly Agree, Agree, \\
        agree with the position you were & Weakly Agree, Weakly Disagree, \\
        tasked with arguing?& Disagree, Strongly Disagree \\
    \hline
        Slightly different:To what extent & Strongly Agree, Agree, \\
        do you honestly agree with the & Weakly Agree, Weakly Disagree, \\
        things that you said in your speech?& Disagree, Strongly Disagree \\
    \hline
        To what extent were you & not at all, a little,\\
        familiar with the issue & informed, well informed\\
        debated in this round? & expert on the issue\\
    \hline
        Regardless of the position you & don't care, little importance \\
        were asked to argue, how & medium important\\
        relevant/important is this &most important of my life,\\
        issue to you personally?& very important\\
    \hline
        How good/strong do you & very poor, poor, fair\\
        think your arguments were?& good, very good\\
    \hline  
        Rank the teams according to how & 1st, 2nd\\
         well you think each team did. &3rd, 4th\\
    \hline
    \end{tabular}
    \label{table:surveyquestion}
\end{table}

\newcommand{\rubriclist}[3]{
    \begin{itemize}[
        nosep, 
        topsep = 0pt,
        partopsep = 0pt,
        leftmargin = *,
        after = \vspace{-\baselineskip},
        before = \vspace{-4pt}
    ]
        \item #1
        \item #2
        \item #3
    \end{itemize}}

\begin{table*}[tbh]
    \centering
    \caption{2019 North American Universities Debate Championships HWS Debate Speaker Judging Point Scale}
    \begin{tabular}{|c|p{6.5in}|}
    \hline
        95-100 &
        \rubriclist
        {Plausibly one of the best debating speeches ever given;}
        {It is incredibly difficult to think up satisfactory responses to any of the arguments made;}
        {Flawless and compelling arguments.}
    \\\hline
        92-94 & 
        \rubriclist
        {An incredible speech, undoubtedly one of the best at the competition;}
        {Successfully engaging with the core issues of the debate, arguments exceptionally well made, and it would take a brilliant set of responses to defeat the arguments;}
        {There are no flaws of any significance.}
    \\\hline
        89-91 & 
        \rubriclist
        {Brilliant arguments successfully engage with the main issues in the round;}
        {Arguments are very well-explained and illustrated, and demand extremely sophisticated responses in order to be defeated;}
        {Only very minor problems, if any, but they do not affect the strength of the claims made.}
    \\\hline
        86-88 &
        \rubriclist
        {Arguments engage with core issues of the debate, and are highly compelling;}
        {No logical gaps, and sophisticated responses required to defeat the arguments;}
        {Only minor flaws in arguments.}
    \\\hline
        83-85 &
        \rubriclist
        {Arguments address the core issues of the debate;}
        {Arguments have strong explanations, which demand a strong response from other speakers in order to defeat the arguments;}
        {May occasionally fail to fully respond to very well-made arguments; but flaws in the speech are limited.}
    \\\hline
        79-82 &
        \rubriclist
        {Arguments are relevant, and address the core issues in the debate;}
        {Arguments well made without obvious logical gaps, and are all well explained;}
        {May be vulnerable to good responses.}
    \\\hline
        76-78 &
        \rubriclist
        {Arguments are almost exclusively relevant, and address most of the core issues;}
        {Occasionally, but not often, arguments may slip into: i) deficits in explanation, ii) simplistic argumentation vulnerable to competent responses or iii) peripheral or irrelevant arguments;}
        {Clear to follow, and thus credit.}
    \\\hline
        73-75 &
        \rubriclist
        {Arguments are almost exclusively relevant, although may fail to address one or more core issues sufficiently;}
        {Arguments are logical, but tend to be simplistic and vulnerable to competent responses;}
        {Clear enough to follow, and thus credit.}
    \\\hline
        70-72 &
        \rubriclist
        {Arguments are frequently relevant;}
        {Arguments have some explanation, but there are regular significant logical gaps;}
        {Sometimes difficult to follow, and thus credit fully.}
    \\\hline
        67-69 &
        \rubriclist
        {Arguments are generally relevant;}
        {Arguments almost all have explanations, but almost all have significant logical gaps;}
        {Sometimes clear, but generally difficult to follow and thus credit the speaker for their material.}
    \\\hline
        64-66 &
        \rubriclist
        {Some arguments made that are relevant;}
        {Arguments generally have explanations, but have significant logical gaps;}
        {Often unclear, which makes it hard to give the speech much credit.}
    \\\hline
        61-63 &
        \rubriclist
        {Some relevant claims, and most will be formulated as arguments;}
        {Arguments have occasional explanations, but these have significant logical gaps;}
        {Frequently unclear and confusing; which makes it hard to give the speech much credit.}
    \\\hline
        58-60 &
        \rubriclist
        {Claims are occasionally relevant;}
        {Claims are not be formulated as arguments, but there may be some suggestion towards an explanation;}
        {Hard to follow, which makes it hard to give the speech much credit.}
    \\\hline
        55-57 &
        \rubriclist
        {One or two marginally relevant claims;}
        {Claims are not formulated as arguments, and are instead just comments.}
        {Hard to follow, almost entirely, which makes it hard to give the speech much credit.}
    \\\hline
        50-55 &
        \rubriclist
        {Content is not relevant;}
        {Content does not go beyond claims, and is both confusing and confused;}
        {Very hard to follow in its entirety, which makes it hard to give the speech any credit.}
    \\\hline
    \end{tabular}
    \label{table:rubric}
    \\[3pt]
    The mark bands above are rough and general descriptions; speeches need not have every future described to fit in a particular band.
    Throughout this scale, 'arguments' refers both to constructive material and responses.
    Judges are asked to use the full range of the scale. 
    Speaker marks determine many of the breaking teams, and tab finishes can be big achievements, so judges are asked to give these marks the serious thought they require.
\end{table*}

\end{document}